\documentclass[9pt,twocolumn,twoside]{opticajnl}

\journal{ol}

\setboolean{shortarticle}{true}

\usepackage{graphicx}
\usepackage{color}
\usepackage{amsmath,bm}

\newcommand{\ket}[1]{\left| #1 \right\rangle}

\setboolean{displaycopyright}{true}

\title{\textcolor{black}{Quantum electrodynamics of chiral and antichiral waveguide arrays}}

\author[1]{Jeremy G. Hoskins}
\author[2]{Manas Rachh}
\author[3*]{John C. Schotland}

\affil[1]{Department of Statistics, University of Chicago, Chicago, Illinois 60637, USA}
\affil[2]{Center for Computational Mathematics, Flatiron Institute, New York, New York 10010, USA}
\affil[3]{Department of Mathematics and Department of Physics, Yale University, New Haven, Connecticut 06511, USA}
\affil[*]{Corresponding author: john.schotland@yale.edu}

\begin{abstract}
\textcolor{black}{
We consider the quantum electrodynamics of single photons in arrays of one-way waveguides, each containing many atoms. We investigate both chiral and antichiral arrays, in which the group velocities of the waveguides are the same or alternate in sign, respectively. We find that in the continuum limit, the one-photon amplitude obeys a Dirac equation. In the chiral case, the Dirac equation is hyperbolic, while in the antichiral case it is elliptic. This distinction has implications for the nature of photon transport in waveguide arrays. Our results are illustrated by numerical simulations.}
\end{abstract}

\setboolean{displaycopyright}{true}

\begin{document}

\maketitle 

There has been considerable recent interest in the quantum electrodynamics of light-matter interactions in waveguide systems~\cite{liao_2016,roy_2017,shen_2005,fang_2015,douglas_2015,goban_2015}. The reduced dimensionality of such systems gives rise to new physical phenomena, especially modifications of spontaneous and stimulated emission. Moreover, the enhanced coupling between atoms and photons can lead to the generation of strong correlations between photons. Similarly, long-ranged interactions between atoms bring about a variety of novel many-body effects. We also note that architectures based on coupled waveguide arrays have emerged as promising candidates for quantum circuits and other quantum technologies.

Waveguides, in which the propagation of light is predominantly in one direction, allow for the intriguing possibility of serving as  one-way carriers of quantum information~\cite{lodahl_2017,pichler_2015,berman_2020}. The attendant violation of reciprocity in such waveguides is due to enhanced spin-orbit coupling of light that is confined at subwavelength scales~\cite{petersen_2014,coles_2016,sollner_2015}. A model of single photons in a one-way waveguide containing a collection of two-level atoms was introduced in \cite{mirza_2017}. It was found that the absence of backscattering prohibits the existence of band structure in periodic systems. Moreover, single-photon transport is unaffected by position disorder, but not by disorder in atomic transition frequencies, where Anderson localization obtains. Related results for three-level atoms and electromagnetically induced transparency have also been described~\cite{mirza_2018}.

In this Letter, we consider the quantum electrodynamics of a single photon in an array of one-way waveguides. The waveguides, each of which contains many two-level atoms, are arranged in a one-dimesional lattice, as illustrated in Fig.~1. Two physical settings are distinguished. \textcolor{black}{For a \emph{chiral} array, the frequencies alternate in value between two interpenetrating sublattices. In an \emph{antichiral} array, the group velocities alternate in sign between the sublattices. In both cases, we will show that in the continuum limit, the one-photon amplitude of a single-excitation state obeys a two-dimensional Dirac equation. In a chiral array, the coordinate along the waveguide is timelike and the transverse coordinate is spacelike. In contrast, in an antichiral array, both coordinates are spacelike. We will see that the distinction between the timelike and spacelike character of the Dirac equations has physical consequences which we illustrate with numerical simulations.}

We begin by considering a one-dimensional array of chiral waveguides. Each waveguide is assumed to contain many two-level atoms. We employ a real-space quantization procedure that treats the atoms and the optical field on the same footing~\cite{mirza_2017}. The Hamiltonian of the system is of the form $H = H_A + H_F + H_I$. Here the atomic Hamiltonian $H_A$ is given by
\begin{equation}
H_A = \omega_0\int dx \sum_n \rho_n(x) \sigma^\dag(x)\sigma(x) \ .
\end{equation}
Here we work in units where $\hbar=1$, $\omega_0$ is the atomic transition frequency of the atoms, and $\rho_n(x)=\sum_j \delta(x-x_{jn})$ is the number density of atoms in the $n$th waveguide, where $x_{jn}$ is the position of atom $j$ in waveguide $n$.
\textcolor{black}{In addition $\sigma(x_{jn})=|0_{jn}\rangle \langle 1_{jn} |$ is the lowering operator for an atom at position $x$ in waveguide $n$, where $|0_{jn}\rangle$ and $|1_{jn} \rangle$ denote the corresponding ground and excited states of the atom.
It follows that $\sigma$ obeys the anticommutation relations
\begin{align}
\label{anticommutation}
\{\sigma(x_{in}),\sigma^\dag(x_{jm})\} = \delta_{ij} \delta_{nm}
\end{align}
and the commutation relations
\begin{align}
[\sigma(x_{in}),\sigma(x_{jm})] = 0 \ ,
\end{align}
with all other commutators and anticommutators vanishing.}
The Hamiltonian of the optical field $H_F$  is of the form
\begin{align}
H_F =  \int dx \sum_n \big[ \phi_n^\dag(x)\left( \Omega_n + i v_n\partial_x \right) \phi_n(x) \\
+ J_0(\phi_n^\dag(x)\phi_{n+1}(x) + \phi_{n+1}^\dag(x)\phi_n(x))\big] \ .
\end{align}
where $v_n$ is the group velocity of the $n$th waveguide and $\Omega_n$ is the frequency about which the waveguide dispersion relation is linearized. In addition, nearest-neighbor waveguides are coupled by evanescent waves, with $J_0$ denoting the corresponding coupling constant.
The field operator $\phi_n(x)$ annihilates a photon at $x$ in waveguide $n$ and obeys the commutation relations
\begin{equation}
\label{commutation}
[\phi_n(x),\phi_m^\dag(y)] = \delta_{nm}\delta(x-y) \ ,
\end{equation}
with all other commutators vanishing.
The interaction between the atoms and the field is described by the Hamiltonian
\begin{align}
H_I =  g \int dx \sum_n & \rho_n(x)\big[\sigma^\dag(x)\phi_n(x) \\
&+ \sigma(x)\phi_n^\dag(x) \big] \ ,
\end{align}
where $g$ is the atom-waveguide coupling constant and we have made the rotating-wave approximation.

We suppose that the system is in a single-excitation state of the form
\begin{align}
\ket{\Psi} =  \int dx \sum_n \big[a_n(x) \sigma^\dag(x)  \\
+ \psi_n(x) \phi_n^\dag(x) \big] \ket{0} \ ,
\end{align}
where $\ket{0}$ is the combined vacuum state of the field and the ground state of the atoms.
Here $a_n(x)$ and $\psi_n(x)$ denote the probability amplitudes of exciting an atom and finding a photon
at the position $x$ in waveguide $n$, respectively. The state $\ket{\Psi}$ obeys the Schrodinger equation $H\ket{\Psi}=E_0\ket{\Psi}$, 
where $E_0$ is the energy. Making use of the Schrodinger equation and the anticommutation relations (\ref{anticommutation}) and commutation relations (\ref{commutation}), we find that 
the $\psi_n$ obey the equations 
\begin{eqnarray}
\nonumber
\label{psi_n}
i  v_n \partial_x\psi_n +\Omega_n\psi_n +  J_0\left(\psi_{n+1}+\psi_{n-1}\right) 
+ V_n(x)\psi_n \\ 
= E_0\psi_n \ ,
\end{eqnarray}
where the potential $V_n$ is defined by
\begin{equation}
V_n(x) = \frac{g^2}{E_0-\omega_0}\rho_n(x) \ .
\end{equation}
We also have that $a_n$ is given by
\begin{equation}
a_n(x) = \frac{ g}{E_0-\omega_0}\rho_n(x) \psi_n(x) \ .
\end{equation}

%%
%\begin{figure}[t]
%\begin{center}
%\textcolor{black}{
%\includegraphics[height=2.3in]{figs/chiral_waveguides.pdf}
%\hspace{15pt}
%\includegraphics[height=2.3in]{figs/antichiral_waveguides.pdf}
%\caption{\textcolor{black}{Illustrating the chiral and antichiral waveguide systems. The chiral case is described
%by the $(1+1)$ dimensional Dirac equation (\ref{dirac_a}). The antichiral waveguide is described
%by the $(2+0)$ dimensional Dirac equation (\ref{dirac_b}).}}}
%\label{waveguides}
%\end{center}
%\end{figure}
%%

\begin{figure}[t]
\begin{center}
\textcolor{black}{
\hspace{-40pt}
\huge Chiral
\vskip 10pt
\includegraphics[height=1.3in]{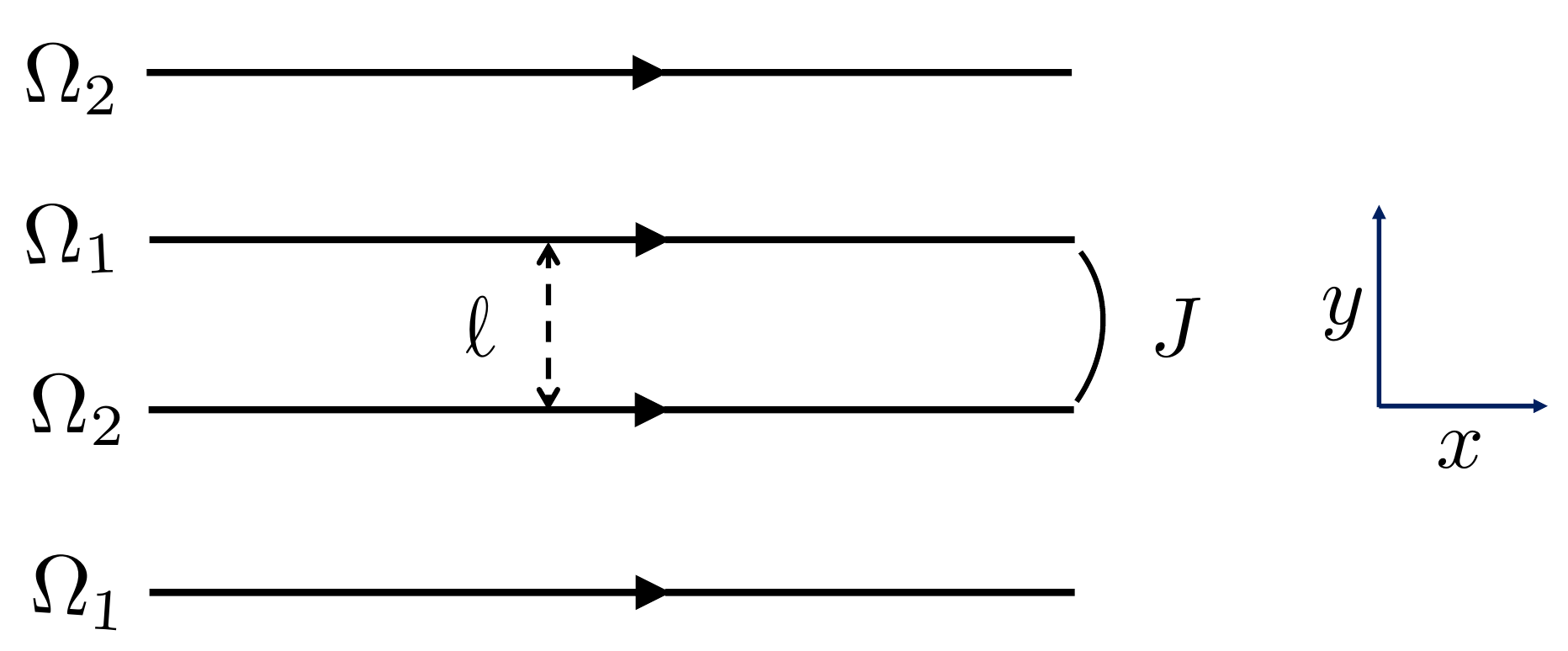}
\vskip .1in
\hspace{-40pt}
\huge Antichiral
\vskip 10pt
\includegraphics[height=1.3in]{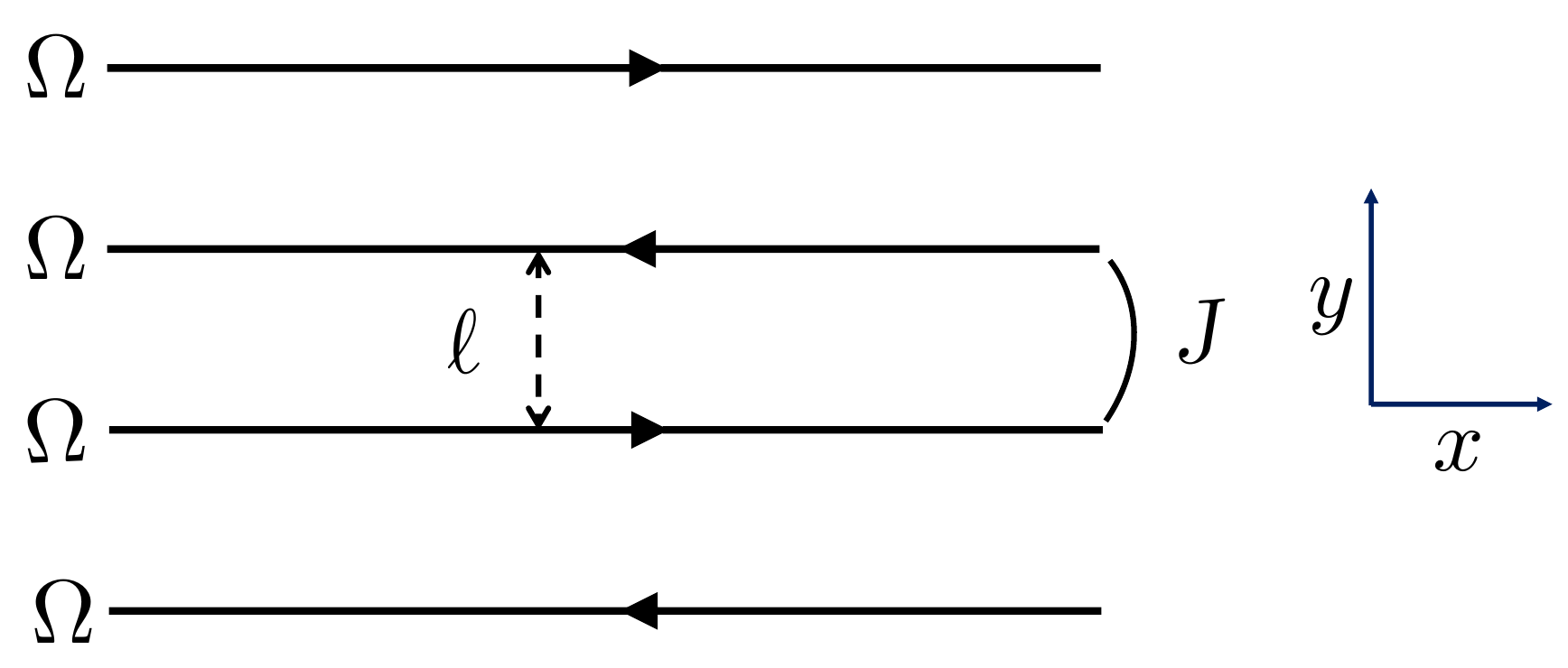}
\caption{\textcolor{black}{Illustrating chiral and antichiral waveguide systems. The chiral case is described
by the $(1+1)$ dimensional Dirac equation (\ref{dirac_a}). The antichiral waveguide is described
by the $(2+0)$ dimensional Dirac equation (\ref{dirac_b}).}}}
\label{waveguides}
\end{center}
\end{figure}

We note an important special case of \eqref{psi_n}. If the waveguides are identical and no atoms are present, then
$V_n=0$, and we can put $v_n=v$ and $\Omega_n=0$. The latter condition can be assumed by appropriately 
shifting the value of $E_0$. ~\eqref{psi_n} then becomes
\begin{eqnarray}
i  v \partial_x\psi_n +  J_0\left(\psi_{n+1}+\psi_{n-1}\right) 
= E_0\psi_n \ ,
\end{eqnarray}
which can be viewed as the equations of motion for a tight binding model~\cite{markos_soukoulis}. It is easily seen that $\psi_n(x)= A \exp(iqx-ik\ell n)$, where $\ell$ is the lattice spacing and the dispersion relation is $E_0=qv + 2J_0\cos(k\ell)$. 

Suppose that the waveguides are arranged as shown in Fig.~1. We define $\tilde\psi_1(x,n)= (-1)^n\psi_{2n}(x)$, $\quad \tilde\psi_2(x,n)=-i (-1)^n \psi_{2n+1}(x)$ and split \eqref{psi_n} into even and odd parts according to
\begin{eqnarray}
\label{lattice_dirac_1}
i v_1 \partial_x \tilde\psi_1 + \Omega\psi_1 + i  J_0\Delta_n \tilde\psi_2 + V\tilde\psi_1= E\tilde\psi_1 \ , \\
i v_2 \partial_x \tilde\psi_2 - \Omega\tilde\psi_2 + i  J_0\Delta_n \tilde\psi_1 + V\tilde\psi_2= E\tilde\psi_2 \ ,
\label{lattice_dirac_2}
\end{eqnarray}
where $v_1$ and $v_2$ are the group velocities in each of the sublattices. 
Likewise, $\Omega_1$ and $\Omega_2$ are the sublattice frequencies, in terms of which we
define $\Omega=(\Omega_{1}-\Omega_{2})/2$ and $E=E_0 + (\Omega_{1}+\Omega_{2})/2$. 
We have also introduced the finite difference operator $\Delta_n f(n) = f(n+1)-f(n)$. 
In the continuum limit, where the spacing $\ell$ between the waveguides vanishes, the ordinary differential equations (\ref{lattice_dirac_1}) and (\ref{lattice_dirac_2}) become the partial differential equations
\begin{eqnarray}
\label{pde_1}
i v_1 \partial_x \psi_1 + \Omega\psi_1 + i  J\partial_y \psi_2 + U(x,y)\psi_1= E\psi_1 \ , \\
i v_2 \partial_x \psi_2 - \Omega\psi_2 + i  J\partial_y \psi_1 + U(x,y)\psi_2= E\psi_2 \ .
\label{pde_2}
\end{eqnarray}
Here $J=J_0\ell$,  $y=n\ell$, $\psi_{1,2}(x,y)=\tilde\psi_{1,2}(x,n)$ and 
\begin{equation}
U(x,y) = \frac{g^2}{E_0-\omega_0}\sum_j\delta(x-x_j)\delta(y-y_j) \ ,
\end{equation}
where $(x_j,y_j)$ is the positions of the $j$th atom. We note that $x$ denotes the coordinate along the waveguide labeled by the transverse coordinate $y$.

Eqs.~(\ref{pde_1}) and (\ref{pde_2}) can be put in the form of a Dirac equation. There are two cases to consider. For a chiral array, the frequencies differ in each sublattice and the group velocities are the same with $v_1=v_2=v$.  We then find that $\psi=(\psi_1,\psi_2)$ obeys the equation
\begin{equation}
\label{dirac_pre}
i v \partial_x\psi + i  J \alpha \partial_y\psi + \Omega \beta \psi + U(x,y) \psi = E\psi \ ,
\end{equation}
where $\alpha$ and $\beta$ are the Pauli matrices
\begin{equation}
\alpha =
\begin{pmatrix} 
      0 & 1 \\
      1 & 0 \\
\end{pmatrix} , \quad 
\beta =
\begin{pmatrix} 
      1 & 0 \\
      0 & -1 \\
\end{pmatrix} \ . 
\end{equation}
It will prove convenient to remove the oscillatory term in \eqref{dirac_pre} by making the transformation $\psi\to e^{-iEx}\psi$. We then see that $\psi$ obeys the Dirac equation
\begin{equation}
\label{dirac_a}
i v \partial_x\psi + i  J \alpha \partial_y\psi + \Omega \beta \psi + U(x,y) \psi = 0 \ .
\end{equation}
\eqref{dirac_a} is a (1+1) dimensional Dirac equation, where the coordinate $x$ is timelike, $y$ is spacelike and $\Omega$ plays the role of the mass. For an antichiral array, the group velocities alternate in sign with $v_1=-v_2=v$ and $\Omega=0$. It follows that $\psi$ obeys
\begin{equation}
\label{dirac_b}
i v \beta \partial_x\psi + i  J \alpha \partial_y\psi  + U(x,y) \psi = E\psi \ .
\end{equation}
\eqref{dirac_b} is a (2+0) dimensional Dirac equation, where both the $x$ and $y$ coordinates are spacelike. We note that similar Dirac equations arise in a variety of physical settings. These include waveguide arrays in classical optics, classical electrodynamics in one dimension, and the one-dimensional tight binding model for binary lattices~\cite{longhi_2010,cannata_1990,shen_2012}.

\textcolor{black}{We now comment on the mathematical and physical differences between the above Dirac equations. In the chiral case, \eqref{dirac_a} is a hyperbolic partial differential equation (PDE). Such equations have wave-like solutions in which singularities propagate. Moreover, the smoothness of the solutions depend on the smoothness of the initial and boundary conditions.  In contrast, in the antichiral case, \eqref{dirac_b} is an elliptic PDE. Equations of this type have smooth solutions, independent of the smoothness of the boundary conditions. }

\begin{figure}[t]
\begin{center}
\textcolor{black}{
\hspace{0.48in} {\huge Chiral} \hspace{1.in}{\huge Antichiral}}

\includegraphics[width=1.6in,height=2.9in]{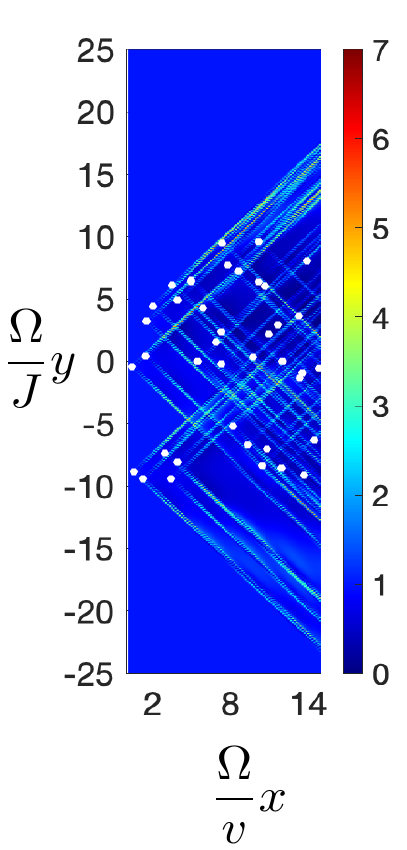}
\hspace{1pt}
\includegraphics[width=1.6in,height=2.8in]{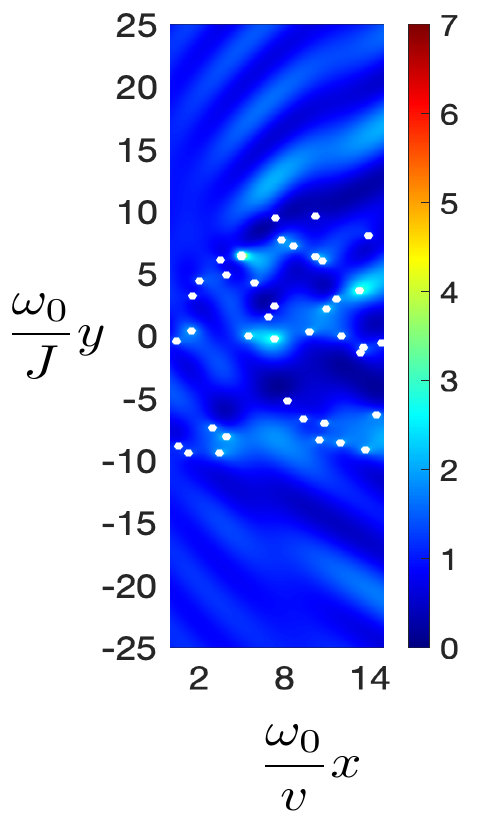}
\caption{\textcolor{black}{Plot of the total probability 
density  $|\psi|^2$ for a single photon in chiral (left) and antichiral (right) waveguide arrays. The atoms are indicated by white dots. The incident field is a plane wave of unit wavenumber propagating in the direction parallel to the waveguides.}}
\label{waveguides}
\end{center}
\end{figure}

\textcolor{black}{We now illustrate the above results with numerical simulations. The Dirac equation~(\ref{dirac_a}) is solved by discretizing the equation in $y$ and then evolving the field in the $x$ direction by solving the resulting system of ordinary differential equations by standard finite difference methods. The Dirac equation~(\ref{dirac_b}) is solved by standard integral equation methods~\cite{carminati_2021}. A fast algorithm is obtained by utilizing recursive skeletonization of the corresponding system of linear algebraic equtions~\cite{Ho_2020}. The field was then calculated using the fast multipole method in time $O(N+M)$, where $N$ is the number of atoms and $M$ is the number of points at which the field is evaluated~\cite{fmm}. In both cases, the incident field is a plane wave of unit wavenumber propagating in the direction parallel to the waveguides. The details are presented in the supplemental materials. }

\begin{figure}[t]
\begin{center}
\huge\textcolor{black}{Periodic}
\includegraphics[height=2.5in]{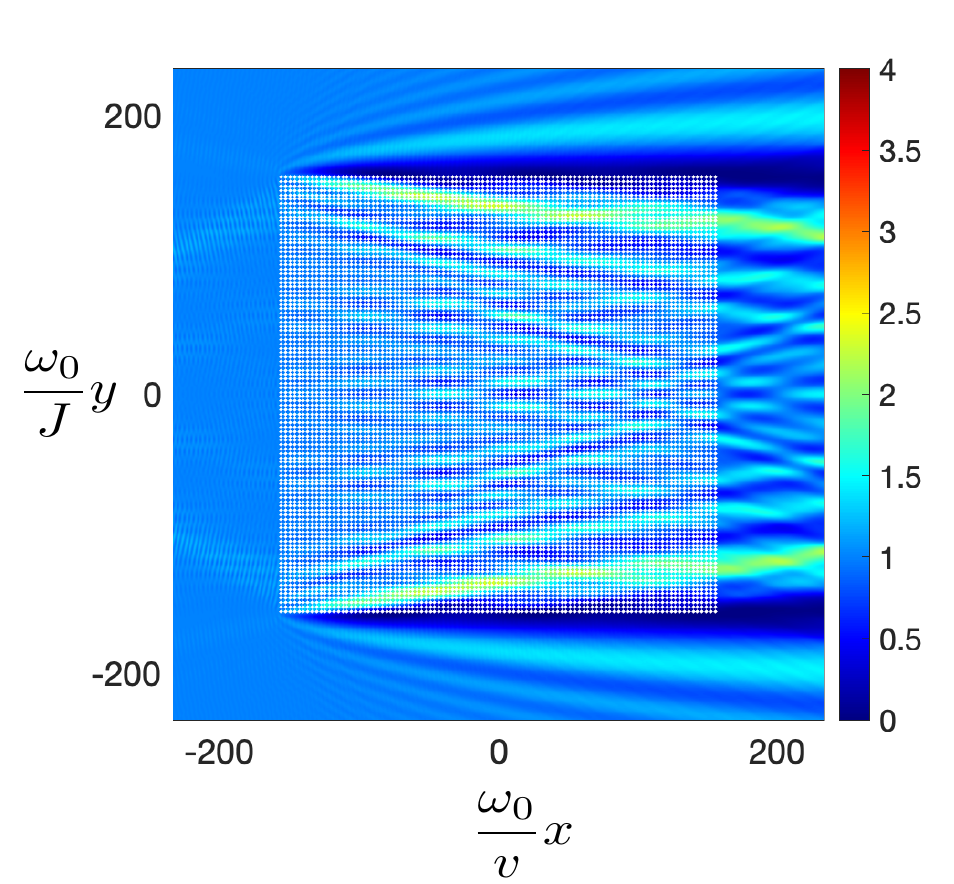}
\vskip 10pt
\huge\textcolor{black}{Random}
\includegraphics[height=2.5in]{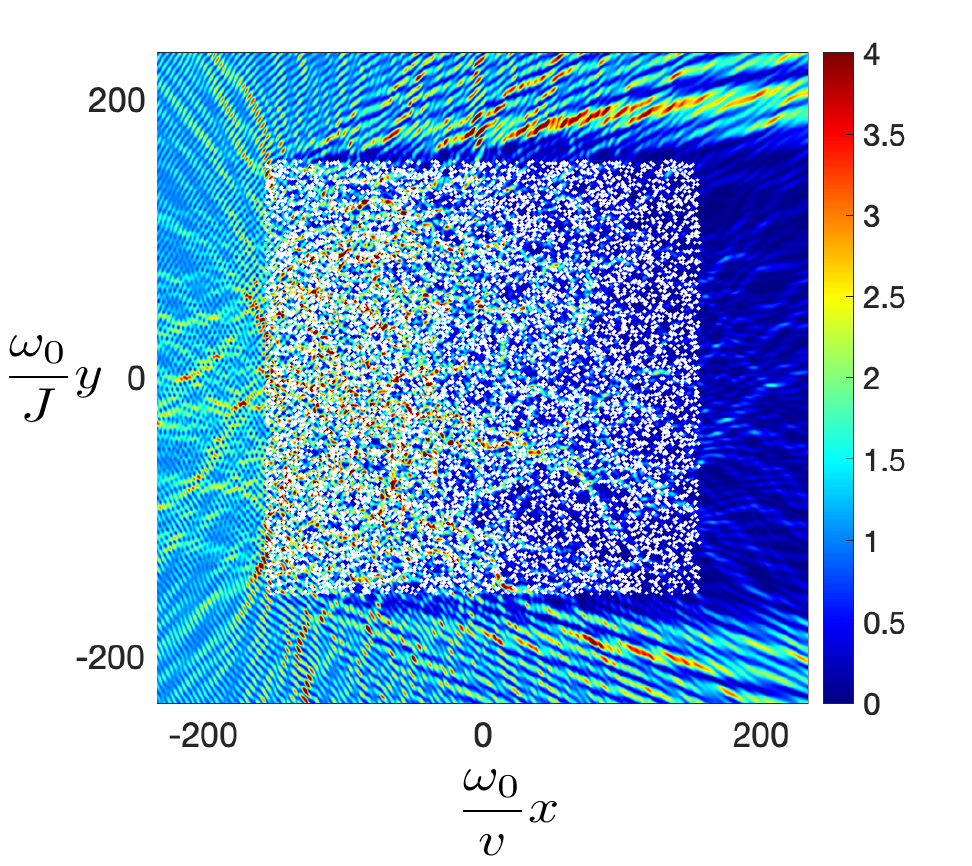}
\caption{\textcolor{black}{Plot of the total probability 
density  $|\psi|^2$ for a single photon in an antichiral waveguide array with $10^4$ atoms.
The atoms are indicated by white dots and arranged periodically (top) and randomly (bottom). The incident field is a plane wave of unit wavenumber propagating in the direction parallel to the waveguides.}}
\end{center}
\vspace{-10pt}
\end{figure}

\textcolor{black}{In Fig.~2 we plot the total probability density $|\psi|^2= |\psi_1|^2 + |\psi_2|^2$ for both chiral and antichiral arrays.
The scattering of a single photon from a collection of 40 atoms is considered. The plot on the left, which corresponds to the chiral case, shows the presence of light cones, consistent with the hyperbolic nature of the Dirac equation~(\ref{dirac_a}). The plot on the right, which corresponds to the antichiral case, demonstrates the smoothing of the wavefronts that is expected from the elliptic nature of the Dirac equation~(\ref{dirac_b}). Evidently, the propagation of light in chiral and antichiral arrays is qualitatively different.}

\textcolor{black}{Fig.~3 further illustrates the case of an antichiral array for a system of $10^4$ atoms. The quantity $|\psi|^2$ is displayed.  In the top plot, the atoms are arranged periodically in a $100\times100$ square lattice. In the bottom plot, the atomic positions are placed randomly, drawn from the uniform distribution on the indicated region.  
We also show results for a system of $10^6$ atoms in Fig.~4., which demonstrates the power of the fast solver.
The physical interpretation of these results requires some care. Diffraction patterns can be observed for both periodic and random systems. We note that in the latter case, a shadow region with boundary rays is visible. This can be analyzed by the usual tools of geometrical optics and diffraction theory applied to the Dirac equation. In the former case, the Dirac equation with a periodic potential can be seen to have a band structure. These results will be presented elsewhere.}

\begin{figure}[t]
\begin{center}
\huge\textcolor{black}{Periodic}
\vskip 5pt
\includegraphics[height=2.5in]{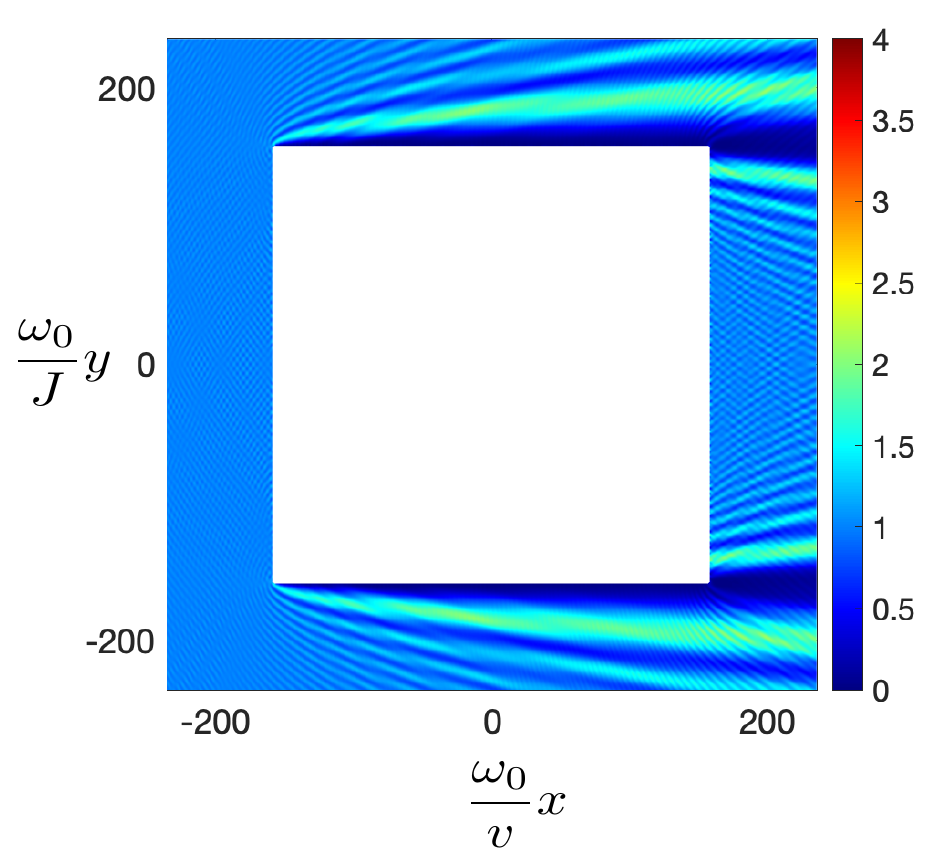}
\vskip 10pt
%\hspace{0pt}
\huge\textcolor{black}{Random}
\vskip 5pt
\includegraphics[height=2.5in]{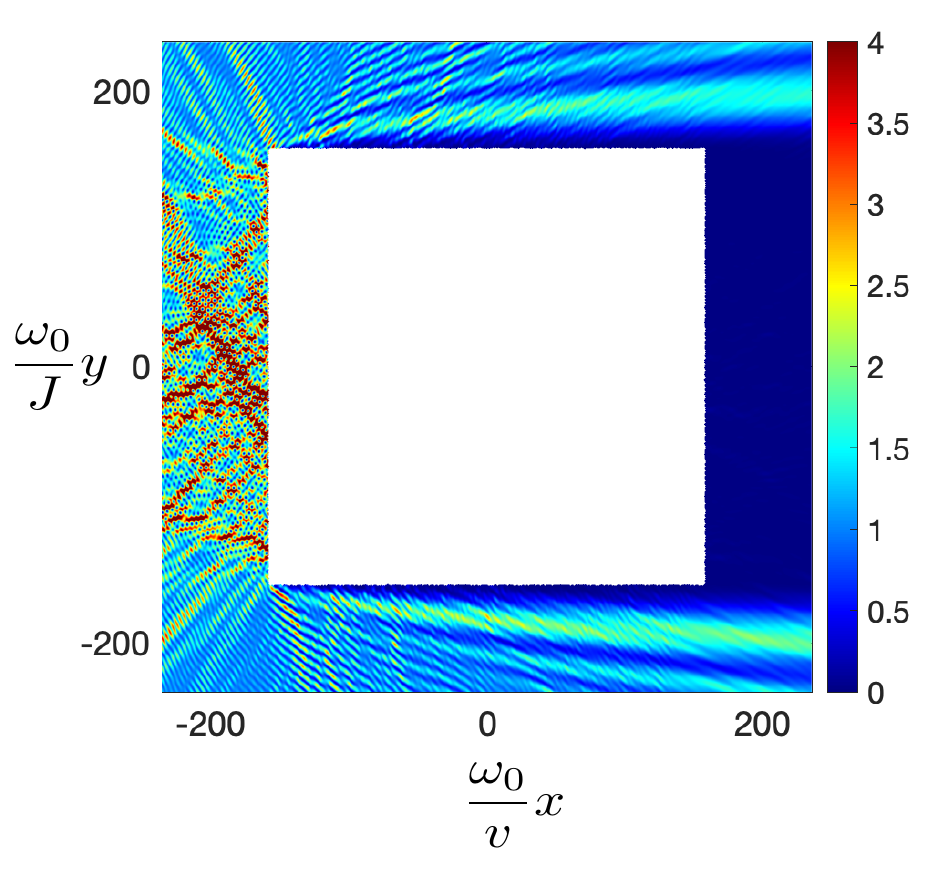}
\caption{Same as Fig.~3  but with $10^6$ atoms.}
\end{center}
\vspace{-10pt}
\end{figure}

\textcolor{black}{
In conclusion, we have investigated the quantum electrodynamics of single photons in arrays of one-way waveguides. We have considered both chiral and antichiral arrays, in which the group velocities of the waveguides are the same or alternate in sign, respectively. We find that in the continuum limit, the one photon amplitude obeys a Dirac equation. In the chiral case, the Dirac equation is  hyperbolic, while in the antichiral case it is elliptic. This distinction has implications for the nature of photon transport in such systems and were demonstrated in numerical simulations. In future work, we plan to investigate the possibility of topological effects in chiral waveguide arrays, as suggested by the presence of a mass term in \eqref{dirac_a}.  In this context, the effects of disorder on the analogs of edge states are of particular interest. It is known that for a single waveguide, there is a subtle interplay between chirality and disorder~\cite{mirza_2017}. Finally, the theory developed in this work has a natural extension to two-photon states in waveguide arrays. In this setting, the transport of entanglement is of particular interest.}

\begin{backmatter}

\bmsection{Funding}
NSF grant DMS-1912821 and AFOSR grant FA9550-19-1-0320.

\bmsection{Disclosures}
The authors declare no conflicts of interest.

\bmsection{Data Availability Statement}
No data were generated or analyzed in the presented research.

\end{backmatter}

\end{document}